\documentclass[aps,prl,reprint,superscriptaddress,footnote, twocolumn,notitlepage]{revtex4-1}

\usepackage{appendix}

\usepackage{tikz}  
\usetikzlibrary{arrows,shapes,positioning,shadows,backgrounds,fit}

\usepackage{relsize}
\usepackage{graphicx}
\usepackage{amsmath, bbm}
\usepackage{amssymb}
\usepackage{amsthm}
\usepackage{mathrsfs}
\usepackage{xcolor}
\usepackage{cancel}
\usepackage{titlesec}
\usepackage{enumitem}  

\usepackage[normalem]{ulem}

\titlespacing{\section}{0ex}{2ex}{0.4ex}
\titleformat{name=\section}{\bf}{\thesection.}{.3em}{}

\usepackage{soul}
\usepackage{dsfont}

\newcommand{\eq}[1]{Eq.\eqref{#1}}

\def\be{\begin{eqnarray}}
\def\ee{\end{eqnarray}}
\newcommand{\tr}[1]{\text{Tr}\left(#1\right)}

\newcommand{\avg}[1]{\left\langle {#1} \right\rangle}

\newcommand{\ii}{\mathcal{I}}
\newcommand{\ww}{\mathcal{W}}

\newcommand{\id}{\mathbb{I}}

\newcommand{\lind}{\mathscr{L}}

\theoremstyle{plain}

\definecolor{myblue}{rgb}{0.2,0.2,0.8}
\definecolor{myblack}{rgb}{0,0,0}
\definecolor{myurl}{rgb}{0.1,0.1,0.4}

\usepackage[colorlinks=true,citecolor=myblue,linkcolor=myblack,urlcolor=myurl]{hyperref}

\begin{document}

\title{Thermodynamic uncertainty relation in slowly driven quantum heat engines} 

\author{Harry J.~D. Miller}
\affiliation{Department of Physics and Astronomy, The University of Manchester, Manchester M13 9PL, UK.}

\author{M. Hamed Mohammady}
\affiliation{RCQI, Institute of Physics, Slovak Academy of Sciences, D\'ubravsk\'a cesta 9, Bratislava 84511, Slovakia}

\author{Mart\'i Perarnau-Llobet}
    \affiliation{D\' epartement de Physique Appliqu\' ee, Universit\' e de Gen\`eve, Gen\`eve, Switzerland}

\author{Giacomo Guarnieri}
\affiliation{School of Physics, Trinity College Dublin, College Green, Dublin 2, Ireland}
\affiliation{Dahlem Center for Complex Quantum Systems,
Freie Universit\"{a}t Berlin, 14195 Berlin, Germany}


\begin{abstract}

Thermodynamic Uncertainty Relations express a trade-off between precision, defined as the noise-to-signal ratio of a generic current, and the amount of associated entropy production. These results have deep consequences for autonomous heat engines operating at steady-state, imposing an upper bound for their efficiency in terms of the power yield and its fluctuations. In the present manuscript we analyze a different class of heat engines, namely those which are operating in the periodic slow-driving regime. 
We show that an alternative TUR is satisfied, which is less restrictive than that of steady-state engines:  it allows for engines that produce finite power, with  small power fluctuations,  to operate close to the Carnot efficiency. 
The bound further incorporates the effect of  quantum fluctuations, which reduces engine efficiency relative to the average power and reliability. We finally illustrate our findings in the experimentally relevant model of a single-ion heat engine.
\end{abstract}

\maketitle

\textit{Introduction:} Much like their macroscopic counterparts, microscopic heat engines function by converting a thermal energy current $J_q$ from their surrounding environment into power $P_w\geq 0$ \cite{Blickle2012,Pekola2015,Ronzani}. In general, such engines can be divided into two classes: steady-state heat engines (SSHEs) and periodically driven heat engines (PDHEs).  SSHEs are comprised of a working substance that is placed in weak contact with multiple reservoirs, so that the ensuing Markovian dynamics results in the engine reaching a non-equilibrium steady-state in the long time limit, thereby supporting a net constant power current~\cite{Benenti}. On the other hand, PDHEs are operated by periodically changing both the mechanical parameters of the working substance, as well as the temperature of its surrounding reservoir, thus generating power by external driving~\cite{Brandner2016a,Mohammady2019b}. In both cases, for any engine operating between a hot  and cold  temperature, $T_h > T_c$, standard thermodynamic laws ensure that the efficiency $\eta := P_w/J_q$ cannot exceed Carnot's bound: $\eta\leq \eta_C := 1-(T_c/T_h)$. In addition to this, microscopic engines are significantly influenced by stochastic fluctuations, which can be of thermal or quantum origin. Understanding how these fluctuations impact the performance of small-scale machines is a central goal of both classical-stochastic~\cite{Seifert2012}, and quantum \cite{campisi2011colloquium,goold2016role},  thermodynamics,  as they determine the engine's reliability.

Recently, Pietzonka and Seifert found that  the efficiency of SSHEs is constrained by a bound tighter than Carnot~\cite{Pietzonka2018}:
\begin{align}\label{eq:seifert}
    \eta\leq \frac{\eta_C}{1+ 2T_c  P_w / \Delta P_w}=: \eta^{PS}.
\end{align}
This bound incorporates an additional dependence on the engine's time-averaged work fluctuations $\Delta P_w$. The quantity $\Delta P_w$ represents the so-called \textit{constancy} of the engine~\cite{Pietzonka2018}, which inversely quantifies the engine's reliability in terms of power output. The bound \eq{eq:seifert} tells us that in order to increase the efficiency of any SSHE, one must either sacrifice the power output $P_w$ or the engine's reliability. This can be seen as a consequence of the \textit{thermodynamic uncertainty relation} (TUR) \cite{Gingrich,Barato,Horowitza}, which states that entropy production constrains the noise-to-signal ratio of any current in SSHEs. Extensions and generalizations of Eq.~\eqref{eq:seifert} to autonomous quantum systems operating at steady-state have been investigated~\cite{Guarnieri2019b, Macieszczak2018, Hasegawa2020PRL, hasegawa2020thermodynamic, Potts2019}.

With regard to PDHEs, it is still currently debated whether a similar universal trade-off is expected to hold: on the one hand, it was found that both in the case of an externally driven Brownian clock~\cite{Barato2016} and in driven cyclic heat engines~\cite{Holubec2018} one can achieve small fluctuations at finite power output in a dissipationless manner. On the other hand, TUR-bounds for driven Langevin systems~\cite{van2020thermodynamic} and dissipative two-level systems~\cite{cangemi2020violation}, as well as for classical time-dependent driven engines~\cite{koyuk2020thermodynamic} were found. 
In general, TURs giving rise to \eq{eq:seifert} can be recovered for protocols that are time-symmetric~\cite{Proesmans}, or modified in order to account for time-asymmetry in the small-amplitude regime~\cite{Macieszczak2018}. Alternatively, other bounds have also been derived with an additional dependence on hysteresis~\cite{Proesmansa,Potts2019} or driving frequency~\cite{Institut2019}. 
However, in all above cases a general \textit{quantum mechanical} trade-off between efficiency, average power and its variance has not yet been achieved. Moreover, the impact of quantum fluctuations on such a trade-off has yet to be established. In this paper, we provide these important missing pieces of the puzzle, by deriving the following quantum version of \eq{eq:seifert} for PDHEs operating in the slow driving, Markovian regime:
\begin{align}\label{eq:eta_bound}
    \eta\leq \frac{\eta_C}{1+ 2T_c P_{w} f\hspace{-0.5mm}\left(\big|\frac{P_\ww}{P_w}\big|\right)   / (\Delta P_w-2\Delta \mathcal{I}_w )}:=\eta^{Q},
\end{align}
where: $P_\ww$ denotes the  adiabatic (also known as \textit{quasi-static}~\cite{Mandal2016a}) power;  $f(x):= (1-x)^2$; and  $0 \leq 2\Delta \mathcal{I}_w \leq \Delta P_w$ is a quantum correction term, which will be precisely defined below.  Firstly, we see that \eq{eq:eta_bound} is structurally different from \eq{eq:seifert} as it now depends on the ratio between actual and adiabatic power. Depending on this ratio, the bound may exceed or fall below the SSHE bound \eq{eq:seifert}. Furthermore, the term $\Delta \mathcal{I}_w$ represents a measure of quantum fluctuations of the power as it depends purely on \textit{quantum friction} \cite{Feldmann,Plastina,Francica2017,Dann2019}, and has recently been shown to lead to a quantum correction to the standard fluctuation-dissipation relation for work \cite{Miller2019,Scandi2019}. Crucially, $\eta^Q$ is a decreasing function with respect to $\Delta \mathcal{I}_w$, meaning that quantum fluctuations have a negative impact on the performance of PDHEs in the slow driving regime, making it impossible to achieve the optimal classical efficiency for a given average power output and its variance.

\textit{Periodic quantum heat engines:} 
We consider engines where the working medium is a driven quantum system, weakly  coupled to a heat bath. Setting $\hbar=k_B=1$, this is described by a time-dependent adiabatic Lindblad master equation of the  form $\dot{\rho}_t=\mathscr{L}_{\lambda(t)}(\rho_t)$ \cite{Albash2012}, where the  time-dependence exhibited by the dynamical generator $\lind_{\lambda(t)}$ is protocol-dependent, and is induced by the external modulation of the bath temperature $T(t)$  and  control mechanical parameters $\Lambda(t)$ which determine the Hamiltonian $H_{\Lambda(t)}$. An engine cycle of duration $\tau$ is then represented by a closed curve in the control parameters space $\lambda : t\mapsto \lambda(t) :=\{T(t),\Lambda(t) \}$, such that it satisfies $\lambda(0)=\lambda(\tau)$.  In particular, following \cite{Brandner2018,Brandner2016a}, we  parameterise the temperature modulation as
\begin{align}\label{tempmodulation}
T(t):=\frac{T_c T_h}{T_h+(T_c-T_h)\alpha(t)}, \ \ T_c \leq T_h,
\end{align}
with $\alpha(t)\in[0,1]$, and $\alpha(0) = \alpha(\tau) = 0$. This implies that at $t=0,\tau$, the  thermal bath that is in contact with the system is at the cold temperature $T_c$, but approaches the hot temperature $T_h$ in the middle of the cycle. From now on we  further assume that, for all~$t$, the quantum detailed balance condition \cite{Alicki2002,Alicki1976} is satisfied, and that there exists a unique stationary state $\pi_{\lambda(t)}$, such that $\lind_{\lambda(t)}(\pi_{\lambda(t)}) =0$,  which is of Gibbs form.   This means that $\pi_{\lambda(t)} = e^{-\beta(t) H_{\Lambda(t)}}/Z_{\lambda(t)}$, where  $\beta(t) := 1/T(t)$ is the inverse temperature and  $Z_{\lambda(t)}:=\tr{e^{-\beta(t) H_{\Lambda(t)}}}$ is the partition function.

A central quantity of interest throughout our analysis is the \textit{non-adiabatic entropy production rate}, defined as
\begin{align}\label{eq:2law}
    \langle \dot \sigma \rangle := \frac{\avg{\sigma} }{\tau}=\frac{1}{\tau} \bigg(\Delta S-\int_0^\tau dt \ \beta(t)\langle \dot{q}(t) \rangle \bigg)\geq 0, 
\end{align}
where $\langle \dot{q}(t) \rangle := \tr{\dot{\rho}_t\, H_{\Lambda(t)}}$ is (in weak coupling) the rate of heat entering the system and $\Delta S$ the increase in information entropy. \eq{eq:2law} quantifies the dissipation in terms of excess heat in order to drive a system out of equilibrium, and can be directly related to the degree of irreversibility of a process \cite{Esposito2010b,Esposito2018}.  
Using \eq{tempmodulation} and the periodic boundary conditions, one can easily show that \eq{eq:2law} takes the form
\begin{align}\label{eq:ent_period}
    \langle \dot{\sigma} \rangle=\frac{1}{T_c}\bigg( \eta_C \, J_q - P_w\bigg) \geq 0,
\end{align}
where we have introduced the time-average power and heat flux supplied to the engine~\cite{Brandner2016a}:
\begin{align}\label{eq:power}
    P_w&:=-\frac{1}{\tau}\int^\tau_0 dt \ \tr{\dot{H}_{\Lambda(t)} \rho_t},\\
    \label{eq:heat}    
    J_q&:=\frac{1}{\tau}\int^\tau_0 dt \ \alpha(t)\tr{ H_{\Lambda(t)} \dot{\rho_t}}.
\end{align}
Naturally, this decomposition leads us to define the efficiency as the ratio $\eta:= P_w/J_q$ between power output and heat flux entering the machine, which for an engine (defined by the regime $P_w\geq 0$) is bounded by the Carnot efficiency $\eta\leq \eta_C := 1-T_c/T_h$ due to the second law \eq{eq:2law}. We note that $\alpha(t)$ plays the role of a weighting function for the heat flux \eq{eq:heat}, with increasing weight assigned to increasing temperatures. This generalises the traditional thermodynamic efficiency where the system interacts with only two baths at distinct temperatures, which is recovered by choosing $\alpha(t)$ to be a step function. In this case, it is easy to see that $J_q$ reduces to the heat flow from the hot bath and the standard definition of efficiency is recovered \cite{Brandner2016a}.

In this paper we are finally concerned with engines that operate in the slow-driving regime, which are characterised by choosing the driving protocol $\lambda(t)$ as a slowly varying periodic function, satisfying boundary conditions $\dot{\lambda}(0)=\dot{\lambda}(\tau)=\{0, 0\}$. This ensures that the system occupies the same equilibrium state $\pi_{\lambda(0)}$ at the start and end of the cycle, and remains close to the  instantaneous steady-state at all times during the cycle, taking the form $\rho_t\simeq\pi_{\lambda(t)}+\delta \rho_t$, where $\delta \rho_t$ is a traceless correction term that vanishes linearly with the driving speed  \cite{Cavina2017,Brandner2020,Miller2019}. This regime is physically reached by setting the engine cycle duration $\tau$ to be large relative to the intrinsic relaxation timescale $t^{eq}$ of the system~\footnote{The intrinsic relaxation timescale $t^{eq}$ of the system is determined by the inverse spectral gap of the Lindbladian.}. In order to evaluate the leading order terms of  Eqs.  \eqref{eq:ent_period} and \eqref{eq:power} in the slow-driving regime, let us first define a self-adjoint operator-valued function  $\delta A_\lambda :  [0, \tau] \ni t \mapsto \delta A_{\lambda(t)}$, where $A_{\lambda(t)}$ can be any self-adjoint operator, and   $\delta:  A \mapsto \delta A := A - \tr{A \pi_{\lambda(t)}}\id$ is a ``spectrum shift'' that ensures $\tr{\delta A_{\lambda(t)}\pi_{\lambda(t)}}=0$ for all $A_{\lambda(t)}$. 
In the Supplemental Material \footnote{See Supplemental Material [url] for mathematical derivations and further details behind the main results, which includes Refs \cite{Fagnola2007,Boullion1971,riechers2018spectral,Fagnola-harmonic-oscillator,Petz2014}}, we show that under the assumption of detailed balance and uniqueness of the instantaneous steady state ($\lind_{\lambda(t)}(\pi_{\lambda(t)}) =0$), the following is a valid inner product for such operator-valued functions:
\begin{align}\label{eq:inner-product}
  &  \nonumber\big<\big< \delta A_\lambda,\delta B_\lambda \big>\big>_\lambda := \frac{1}{2 \tau} \int_0^\tau dt \int^\infty_0d\theta \\
    &  \tr{\delta A_{\lambda(t)}(\theta) \ \mathbb{J}_{\lambda(t)} ( \delta B_{\lambda(t)}) +  \delta B_{\lambda(t)}(\theta) \ \mathbb{J}_{\lambda(t)} (\delta A_{\lambda(t)})},
\end{align}
where  $\delta A_{\lambda(t)}(\theta) := e^{\theta \mathscr{L}_{\lambda(t)}^*}(\delta A_{\lambda(t)})$, with $\lind_{\lambda(t)}^*$ the generator in the Heisenberg picture, and $\mathbb{J}_{\lambda(t)}(\cdot):=\int^1_0 ds \ \pi_{\lambda(t)}^s (\cdot) \pi_{\lambda(t)}^{1-s}$.
The results of \cite{Scandi,Cavina2017,Miller2019,Brandner2020,Abiuso2020} provide a means of Taylor expanding Eqs.  \eqref{eq:ent_period} and \eqref{eq:power} up to first order in $t^{eq}/\tau$, which can be shown in terms of the inner product \eq{eq:inner-product} as 
\begin{align}
& \langle \dot \sigma \rangle =  \big<\big<   \delta \dot{X}_\lambda,  \delta \dot{X}_\lambda \big>\big>_\lambda, &P_{ w} = P_\ww - \big<\big<  \delta \dot{H}_\lambda,   \delta \dot{X}_\lambda \big>\big>_\lambda , \label{eq:inner_av}
\end{align}
where $X_{\lambda(t)} := \beta(t) H_{\Lambda(t)}$, while 
$\delta \dot{H}_{\lambda(t)} \equiv  \delta \dot{H}_{\Lambda(t)}$. 
Moreover, we have  introduced the \textit{adiabatic power} as 
\begin{align}
\label{eq:adwork}
    P_\ww := -\frac{1}{\tau} \int_0^\tau dt \  \tr{\dot{H}_{\Lambda(t)}\pi_{\lambda(t)}},
\end{align}
which is the engine's power  assuming that the system is in equilibrium at all times (compare with \eq{eq:power}), achieved in the limit $t^{eq}/\tau \to 0$. 

So far we have only considered ensemble averages of thermodynamic quantities. For quantum-mechanical systems, the higher order statistics associated with work become preponderant and fundamentally depend on the measurement scheme used to monitor the system. In the case of open quantum systems whose dynamics are described by a time-dependent Lindblad master equation, the fluctuating work can be determined at the stochastic level by monitoring sequences of quantum jumps exhibited by the system as it interacts with an environment~\cite{Horowitz2012,Horowitz2013b,Horowitz2014,Manzano2015,Liu2016a,FeiLiu2,Manzano2018,Elouard2018a,Mohammady2019d,Menczel2020}. Each time a jump occurs heat is exchanged between engine and environment, and this can be experimentally monitored through an external quantum detector~\cite{Murch2013,Pekola2013b,Naghiloo2019}.  Alternatively, one may determine the fluctuating work from two-time global energy measurements, on both the system and bath, at the beginning and end of the cycle~\cite{Talkner2007,Esposito2009}. In the Markovian limit with weak coupling between system and bath, both approaches allow one to arrive at a general expression for the time-averaged work variance, dependent only on the system degrees of freedom \cite{Miller2019,Miller2020c}, which takes the following form in the slow driving limit:
\begin{align}
     \Delta P_w &= 2\left(\Delta \mathcal{I}_w+ \big<\big<  \delta \dot{H}_\lambda,  \delta \dot{H}_\lambda \big>\big>_\lambda \right).\label{eq:inner_av3}
\end{align}
Here, we have identified  a quantum correction term due to the fluctuations,
\begin{align}\label{eq:friction}
    \Delta \mathcal{I}_w:= \frac{1}{\tau} \int^\tau_0 dt \ \tau^{eq}_{\lambda(t)} \ \mathcal{I}_{\lambda(t)}(\dot{H}_{\Lambda(t)},\dot{H}_{\Lambda(t)}),
\end{align}
 where we introduce the \textit{skew covariance} \cite{DenesPetz2009,Hansen2008}
\begin{align}\label{eq:skew_cov}
    \mathcal{I}_{\lambda(t)}(A,B):=-\frac{1}{2}\int^1_0 ds \ \tr{[A,\pi_{\lambda(t)}^s][B,\pi_{\lambda(t)}^{1-s}]}.
\end{align}
The skew information $\mathcal{I}_{\lambda(t)}(A,A)\geq 0$ represents a measure of quantum fluctuations in the sharp observable $A=A^\dagger$ with respect to instantaneous equilibrium $\pi_{\lambda(t)}$ \cite{Wigner1963a,Luo2006,Marvian2014a,Shitara2016,Frerot2016}. In particular, the skew information vanishes for $[A,\pi_{\lambda(t)}]=0$, reduces to the usual variance $\langle A^2 \rangle-\langle A \rangle^2$ for pure states, and is convex under classical mixing. In this context, $\mathcal{I}_{\lambda(t)}(\dot{H}_{\Lambda(t)},\dot{H}_{\Lambda(t)})$ measures the degree of quantum power fluctuations due to the generation of \textit{quantum friction} stemming from $[\dot{H}_{\Lambda(t)},H_{\Lambda(t)}]\neq 0$ \cite{Feldmann,Plastina,Francica2017,Dann2019,Miller2019,Scandi2019}. Additionally, these quantum fluctuations are weighted by an integral relaxation timescale:
\begin{align}\label{eq:relax}
    \tau^{eq}_{\lambda(t)}:=\int^\infty_0 d\theta \ \frac{\mathcal{I}_{\lambda(t)}(\dot{H}_{\Lambda(t)}(\theta),\dot{H}_{\Lambda(t)}(0))}{\mathcal{I}_{\lambda(t)}(\dot{H}_{\Lambda(t)}(0),\dot{H}_{\Lambda(t)}(0))}\geq 0.
\end{align}
This quantifies the timescale over which the quantum correlation function for the power decays to its equilibrium value, and can be viewed as a quantum generalisation of the integral relaxation time employed in classical non-equilibrium thermodynamics \cite{Feldmann2013,Sivak2012a,Mandal2016a}.

\begin{figure*}[htbp!]
\begin{center}
\includegraphics[width=0.9\textwidth]{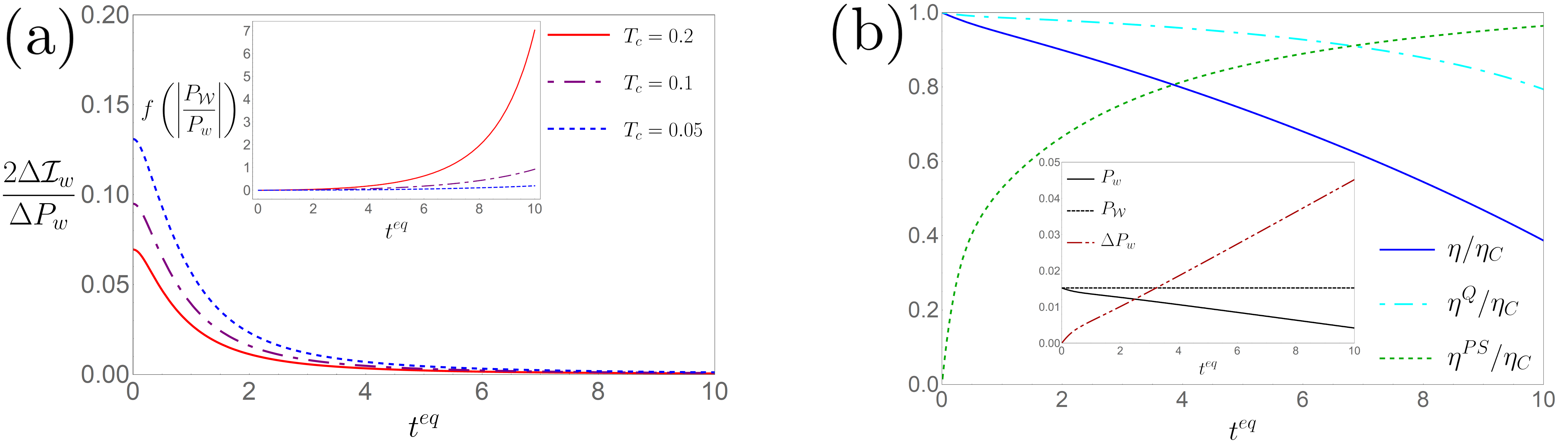}
\vspace*{-0.2cm}
\caption{
Here, we simulate the single ion engine, defined by the protocol in \eq{eq:protocol}, with the parameters $\omega_0=1$,  $T_h=2$, and $\tau = 100$, where we choose units of $\hbar = k_B = 1$. (a)  Plot of the ratio between quantum and total power fluctuations, $2\Delta \ii_w/ \Delta P_w$, as a function of $t^{eq} :=1/\Gamma$ and  $T_c$.  In the inset, we evaluate the correction term for the bound \eq{eq:eta_bound} due to the ratio between the adiabatic and actual power. (b)  Plot of the efficiency $\eta$ and the bounds $\eta^{PS}$, \eq{eq:seifert},  and $\eta^Q$, \eq{eq:eta_bound},  as a function of $t^{eq}:= 1/\Gamma$, for  $T_c = 0.2$. In the inset, we evaluate the power $P_w$, adiabatic power $P_\ww$, and power fluctuations $\Delta P_w$.}
\vspace*{-0.5cm}
\label{fig:bound}
\end{center}
\end{figure*}

\textit{Quantum bound on efficiency.} We are now ready to derive  bounds on the performance of quantum PDHEs. By noting that  $\langle \dot{\sigma} \rangle$, $P_w$, and $ ( \Delta P_w -2 \Delta \mathcal{I}_w)$ can all be expressed through the inner product introduced in \eq{eq:inner-product}  
we can apply the Cauchy-Schwarz inequality $\big<\big< A_\lambda,A_\lambda\big>\big>_\lambda  \big<\big< B_\lambda,B_\lambda\big>\big>_\lambda  \geq |\big<\big< A_\lambda,B_\lambda\big>\big>_\lambda|^2 $, thus obtaining our central result:
\begin{align}\label{eq:TUR}
    ( \Delta P_w -2 \Delta \mathcal{I}_w)\langle \dot{\sigma} \rangle\geq  2 f\left(\left|\frac{P_\ww}{P_w} \right|\right) P_w^2,
\end{align}
where $f(x):= (1- x)^2$. The above inequality is a quantum generalisation of the TUR for entropy production and  power in PDHEs. It demonstrates a trade-off between the entropy production rate $\langle \dot{\sigma} \rangle$ and the noise-to-signal ratio of  power, $\sqrt{\Delta P_w}/P_{ w}$. This TUR has an immediate consequence for the achievable engine efficiency:  a simple rewriting of \eq{eq:ent_period} as $\eta = \eta_C P_w/ (T_c \avg{\dot \sigma} + P_w)$, and combination with \eq{eq:TUR} produces the desired efficiency bound \eq{eq:eta_bound}.

When comparing \eq{eq:eta_bound}  to the  TUR bound for SSHEs  \eq{eq:seifert}, we notice a modification  stemming both from the ratio between adiabatic and actual power $|P_\mathcal{W}/P_w|$, as well as from the presence of $\Delta \mathcal{I}_w$.   We show in the Supplemental Material \footnotemark[2] that this quantum correction satisfies $0 \leq 2\Delta \mathcal{I}_w \leq \Delta P_w$, which means  that  $\eta^{Q}$ decreases monotonically with increasing $\Delta \mathcal{I}_w$. As a consequence, the bound   \eq{eq:eta_bound} is in fact more restrictive than the equivalent classical engine bound with vanishing quantum fluctuations, namely $\eta^Q \leq \eta^\mathrm{cl} :=  \eta^Q  \text{ s.t. } \Delta \mathcal{I}_w = 0$. This demonstrates the detrimental influence of quantum friction for PDHEs close to equilibrium. Indeed, to saturate \eq{eq:eta_bound} we require that, for all times $t \in [0,\tau]$,  $\delta \dot{H}_{\lambda(t)} \propto \delta \dot{X}_{\lambda(t)}$. However, this necessarily implies a vanishing quantum friction $[\dot{H}_{\Lambda(t)},H_{\Lambda(t)}]=0$ for all $t\in [0,\tau]$, which is a necessary and sufficient condition for  $\Delta \mathcal{I}_w=0$. Hence, \eq{eq:eta_bound} is in fact a strict inequality for heat engines in the presence of quantum friction. As expected and discussed in Refs.~\cite{Miller2019,Scandi2019}, the quantum correction becomes most relevant at low temperatures, which we will later illustrate by an example. 

The correction arising from $|P_\mathcal{W}/P_w|$ affects engines irrespectively of quantum friction, and can  lead to large deviations from \eq{eq:seifert}. First, we note that for time-symmetric protocols, where $P_\mathcal{W}=0$, we recover the original bound, which is in agreement with \cite{Proesmans}.  Moving slightly away from this point, one may find a regime where $P_\mathcal{W}>2P_w$. In this case, $f(|P_\mathcal{W}/P_w|)>1$, and the bound \eq{eq:eta_bound} becomes more restrictive than that of SSHEs  \eq{eq:seifert}. However, in most operating regimes with non-vanishing power output we expect $f(|P_\mathcal{W}/P_w|)\ll 1$, as $P_\mathcal{W}-P_w = \mathcal{O}(t^{eq}/\tau) \ll 1$ in the slow driving regime, see the expansion \eq{eq:inner_av} (recall that $t^{eq}$ is the characteristic equilibration timescale).  That is, the power-efficiency of slowly driven PDHEs is  less constrained by fluctuations than that of SSHEs. To quantify this, one may expand \eq{eq:eta_bound} in $\epsilon\equiv t^{eq}/\tau$ obtaining:
\begin{align}
    \eta \leq \eta_C \left(1- \epsilon \frac{ 2T_c    a_P^2}{P_\mathcal{W} a_{\Delta P}}  \right)+\mathcal{O}\left( \epsilon^2 \right),
    \label{etaexp}
\end{align}
where  we have defined $a_P \equiv \lim_{\epsilon \rightarrow 0} (P_w -P_\mathcal{W})/\epsilon$ and $a_{\Delta P} \equiv \lim_{\epsilon \rightarrow 0} (\Delta P_w-2\Delta \ii_w)/\epsilon$, which are finite and can be inferred from Eqs. \eqref{eq:inner_av} and \eqref{eq:inner_av3}. 

Since the bound can be saturated for $\delta \dot{H}_\lambda \propto \delta \dot{X}_\lambda$, one can in principle approach the Carnot efficiency, at finite power and zero work fluctuations, in the limit where  $t^{eq}$ becomes vanishingly small and $\tau$ remains finite (recall that work fluctuations are proportional to $\epsilon$ from \eq{eq:inner_av3}). This is  in accordance with the results demonstrated in \cite{Holubec2018,Denzler2020}. Our bound \eq{eq:eta_bound} thus incorporates previous results and furthermore gives the leading order correction to Carnot's bound for finite speed  and equilibration time-scale, while clarifying that PDHEs also obey TUR relations.

\textit{Single ion heat engine:} To illustrate our bound \eq{eq:eta_bound} we consider a model of a single ion PDHE, inspired by recent experimental realisations using ion-traps \cite{Rossnagel2015}. We describe the engine using a master equation for the damped harmonic oscillator:
\begin{align}
    \dot{\rho}_t=- i \omega [a_\omega^{\dagger} a_\omega, \rho_t]+\Gamma(N_\beta+1) \mathcal{D}_{a_\omega}[\rho_t]+\Gamma \mathcal{D}_{a_\omega^{\dagger}}[\rho_t],
    \label{eq:mastereq}
\end{align}
with $\mathcal{D}_X[\rho]= X\rho X^{\dagger}- \frac{1}{2}\{X^{\dagger}X,\rho\}$. Here the Hamiltonian is $H_\omega=\omega (a^\dagger_\omega a_\omega+\frac{1}{2})$ with $\omega$ the time-dependent frequency, $a_\omega=\sqrt{\omega/2}(x+i p/\omega)$ is the creation operator with unit mass, $\Gamma$ is the damping rate (in the slow driving regime,  $t_{\rm eq}/ \tau \ll 1 $ with $t_{\rm eq}\equiv \Gamma^{-1}$), and $N_\beta=1/(e^{\beta \omega}-1)$ is the Bose-Einstein distribution. We consider a cycle defined by the  slow modulation of the engine's oscillator frequency and bath temperature,  $\lambda:t\mapsto \lambda(t)=\{T(t),\omega(t) \}$,  according to the periodic functions
 \begin{align}\label{eq:protocol}
     &\omega(t) = \omega_0 \left(1+\frac{1}{2}\sin\left(\frac{2\pi t}{\tau}\right)+\frac{1}{4}\sin\left(\frac{4\pi t}{\tau} + \pi\right) \right),
     \nonumber\\
     & T(t)= \frac{T_c T_h}{T_h+(T_c-T_h)\sin^2 \left(\frac{\pi t}{\tau} \right)} ,
\end{align}
where $T_c < T_h$ and $\omega_0 >0$. Note that for the temperature, we have simply assigned $\alpha(t) = \sin^2(\pi t /\tau)$ in \eq{tempmodulation}.  This protocol is cyclic, $\omega(0) = \omega(\tau) = \omega_0, T(0) = T(\tau) = T_c$, and satisfies the slow-driving condition $\dot{\omega}(t) = \dot{T}(t) = 0$ for $t = 0, \tau$.  In the Supplemental Material \footnotemark[2] we calculate  the  power and its fluctuations, as well as the efficiency and its bounds, using Eqs. \eqref{eq:ent_period}, \eqref{eq:inner_av},  and \eqref{eq:inner_av3}. Notably, the power operator $\dot{H}_\omega=\dot{\omega}\big(\omega^{-1}H_\omega+((a^\dagger_\omega)^2+a_\omega^2)/2\big)$ does not commute with the engine Hamiltonian, $[H_\omega,\dot{H}_\omega]\neq 0$, meaning that quantum friction is present throughout the cycle, and so the quantum correction term \eq{eq:friction} is strictly positive. To see this effect, we plot the ratio between quantum and total power fluctuations, i.e. $2\Delta \ii_w/\Delta P_w$, in Fig.~\ref{fig:bound} (a). It can be seen how the quantum fluctuations become more relevant in the low temperature regime,  $\omega_0/T_c \gg 1$, as expected.  In this regime, the TUR \eq{eq:TUR} might become substantially affected by quantum fluctuations. Moreover, while the total power fluctuations $\Delta P_w$ vanish as $t^{eq}/\tau \to 0$ (see the inset of Fig.~\ref{fig:bound} (b)), the ratio of the quantum fluctuations $2\Delta \ii_w/\Delta P_w$ can be seen to increase as $t^{eq}/\tau$ becomes smaller, showing that  quantum fluctuations becomes more relevant in the slow-driving limit.   Conversely, the correction term $f(|P_\mathcal{W}/P_w|)$ is large when $t^{eq}/\tau$ is large, and vanishes in the limit $t^{eq}/\tau \to 0$.  This means that, as can be seen in  Fig.~\ref{fig:bound} (b), in the limit $t^{eq}/\tau \to 0$ the engine produces finite power $P_w = P_\ww$, while both the efficiency $\eta$, as well as the bound  given in \eq{eq:eta_bound}, approaches Carnot. Finally, we see in  Fig.~\ref{fig:bound} (b) that the bound in \eq{eq:seifert} does not apply to PDHEs; while the efficiency $\eta$ always obeys \eq{eq:eta_bound}, it can violate \eq{eq:seifert} for sufficiently small $t^{eq}/\tau$, as this bound vanishes in the slow-driving limit, due to the fact that  $\Delta P_w$ becomes vanishingly small.

 \textit{Conclusions:} We have derived a bound on the optimal efficiency of quantum periodically driven heat engines (PDHEs) in terms of their average power and constancy, valid in the slow-driving, Markovian regime. In the first instance, we see that PDHEs are subject to a bound that differs from steady-state heat engines (SSHEs) through an additional dependence on the ratio between adiabatic and actual power. Nonetheless, \eq{eq:eta_bound} still imposes a universal constraint on engine efficiency for a given power and constancy, thus providing a finite time correction to the Carnot bound at leading order in driving speed. The bound further incorporates the effect of quantum friction stemming from possibly non-commuting Hamiltonian driving. This represents the first Thermodynamic Uncertainty Relation  for PDHEs that explicitly shows the role of  quantum effects.

It has recently been shown that quantum friction reduces the maximum power achievable in slow driving PDHEs~\cite{Brandner,Brandner2020,Abiuso2020,Abiuso2020II}. Our results demonstrate that in this operational regime, quantum friction limits the efficiency relative to the subsequent reliability and power. More specifically, when optimising any one of the trio $\phi\in\{\eta, P_w, 1/\Delta P_w\}$ while fixing the other two variables, quantum friction inevitably leads to a reduction in the maximum value $\phi_{\max}$ that can be attained. 
Given that enhancements with a  quantum origin have been identified in other thermodynamic contexts, such as Otto-like engines  \cite{Uzdin2015,Uzdin2019,lostaglio2020certifying} or refrigerators \cite{Brunner2014},  a full understanding of the role of quantum effects in PDHEs beyond the slow-driving and weak-coupling regime remain as open questions. 

\begin{acknowledgments}
\emph{Acknowledgements:} H. J. D. M. acknowledges support from the Royal Commission for the Exhibition of 1851. M. H. M. acknowledges support from  the Slovak Academy of Sciences   under MoRePro project OPEQ (19MRP0027), as well as projects OPTIQUTE (APVV-18-0518) and HOQIT (VEGA 2/0161/19).  M. P.-L. acknowledges funding from Swiss National Science Foundation (Ambizione  PZ00P2-186067). G. G. acknowledges
fundings from FQXi and DFG Grant No. FOR2724. G. G.
also acknowledges funding from European Unions Horizon
2020 research and innovation programme under the Marie
Sklodowska-Curie Grant Agreement No. 101026667.
\end{acknowledgments}

\bibliographystyle{apsrev4-1}
\bibliography{mybib}




\end{document}